\documentclass[pre,twocolumn,showpacs,preprintnumbers,amsmath,amssymb]{revtex4}
\usepackage{dcolumn}
\usepackage{bm}
\usepackage{amsmath,graphicx}
\usepackage{color}

\newcommand{\ep}{\varepsilon}

\begin{document}
\title{Curvature effects on collective excitations in dumbbell-shaped hollow nanotubes}

\author{Hiroyuki Shima}
\email[Email-address:]{shima@eng.hokudai.ac.jp}
\affiliation{Department of Applied Physics, Graduate School of Engineering,
Hokkaido University, Sapporo 060-8628, Japan}

\author{Hideo Yoshioka}
\affiliation{Department of Physics, Nara Women's University, Nara 630-8506, Japan}

\author{Jun Onoe}
\affiliation{Research Laboratory for Nuclear Reactors and Department of Nuclear Engineering,
Tokyo Institute of Technology, 2-12-1 Ookayama, Meguro, Tokyo 152-8550, Japan}

\date{\today}

%
%

\begin{abstract}
We investigate surface-curvature induced alteration in the Tomonaga-Luttinger liquid (TLL) states of a one-dimensional (1D) deformed hollow nanotube with a dumbbell-shape. Periodic variation of the surface curvature along the axial direction is found to enhance the TLL exponent significantly, which is attributed to an effective potential field that acts low-energy electrons moving on the curved surface. The present results accounts for the experimental observation of the TLL properties of 1D metallic peanut-shaped fullerene polymers whose enveloping surface is assumed to be a dumbbell-shaped hollow tube.
\end{abstract}



\maketitle

\section{Introduction}

When the motion of a quantum particle is constrained to a geometrically curved surface,
the surface curvature produces an effective potential field
that affects spatial distribution of the wavefunction amplitude \cite{daCosta}.
Such the geometric curvature effect on quantum states
has been discussed in the early-stage development of the quantum mechanics theory \cite{DeWitt,schuster}.
Still in the last years,
the effect has focused renewed attention in the field of condensed matter physics,
mainly due to technological progress that enables to fabricate 
low-dimensional nanostructures with complex geometry \cite{prinz,mci,terrons,twist1,ntn,twist2,pssa}.
Several intriguing phenomena as to single-electron transport in curved geometry
have been theoretically predicted \cite{exp1,exp2,exp3,exp4,exp5,exp6,exp7,exp8,exp9,QHE,shota},
in addition to curvature-induced anomalies in classical spin systems \cite{spin1,spin2,spin3,spin4,spin5}
which imply untouched properties of quantum counterparts.
Besides curvature ones, geometric torsion effects on quantum transport of
relativistic \cite{jensen} and non-relativistic \cite{taira} particles 
through twisted internal structures have been also discussed very recently.

\begin{figure}[bbb]
\begin{center}
\includegraphics[scale=0.35]{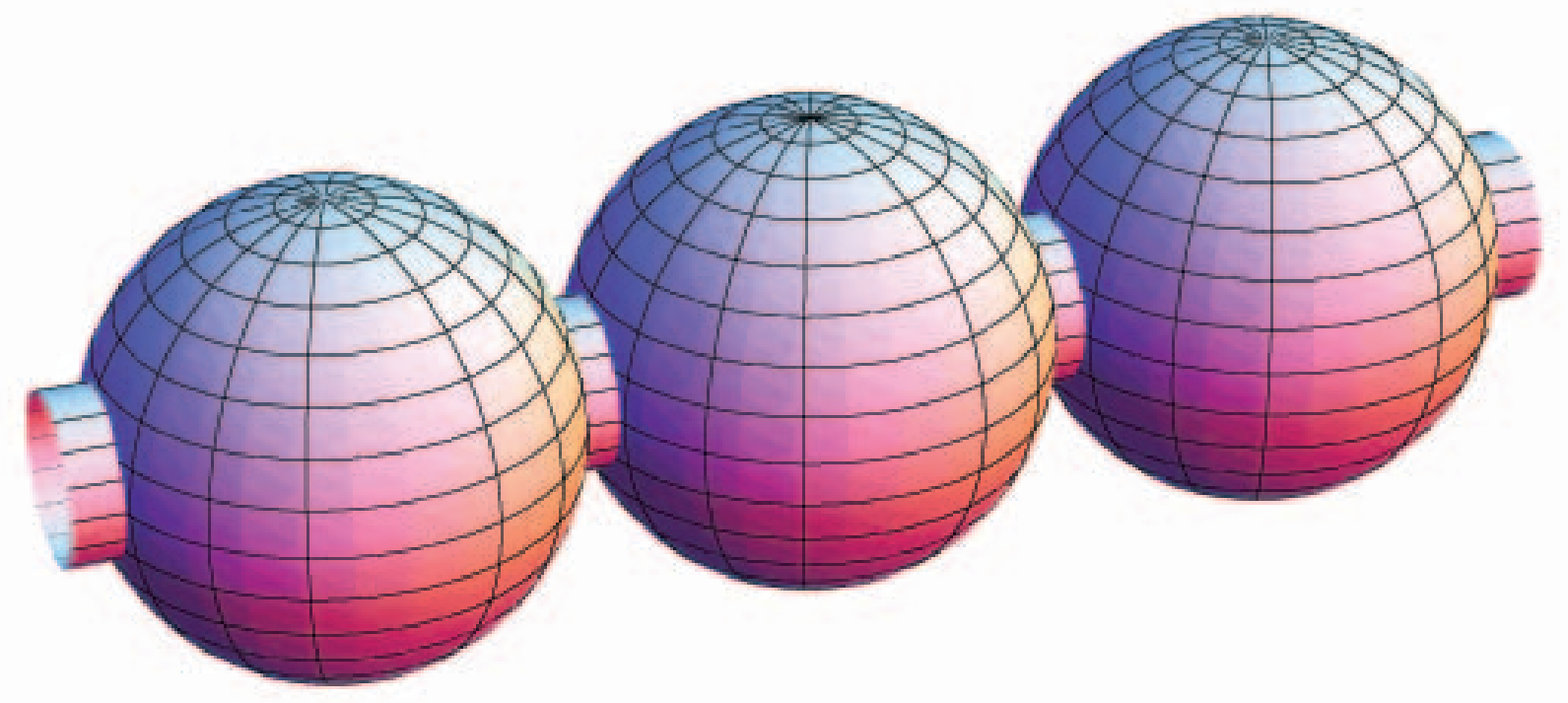}
\includegraphics[scale=0.33]{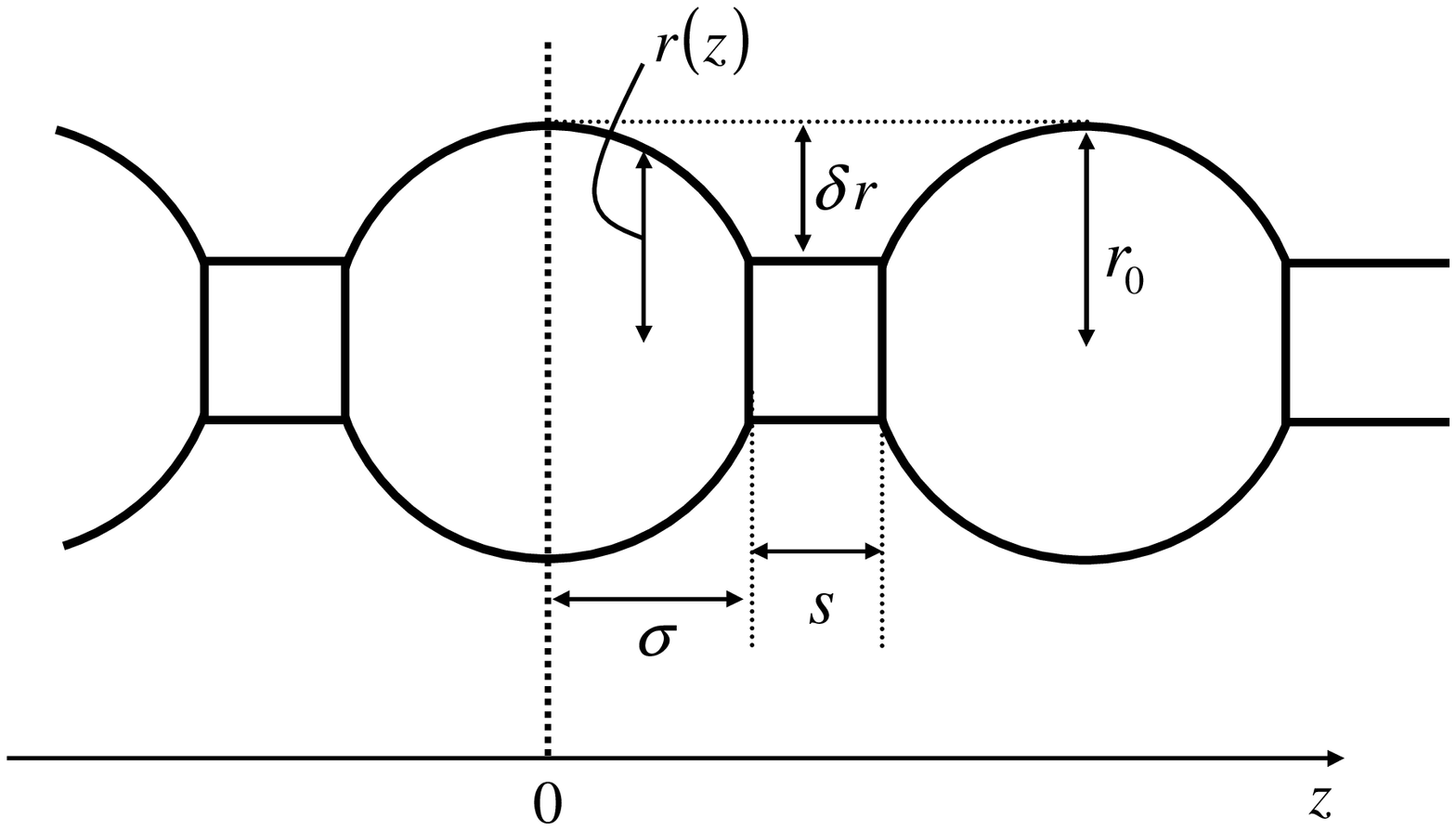}
\end{center}
\caption{(color online) Top: Schematic illustration of a dumbbell-shaped hollow nanotube.
It consists of a chain of large spherical surfaces threaded by a thin hollow tube.
Bottom: Cross section of the dumbbell-shaped hollow tube along with the tube axis $z$.
The neck length $s$ serves as the unit of length, and the radius of spheres $r_0/s = 4$ is fixed 
throughout calculations.
The width $\sigma$ of spherical regions is determined by $\delta r$ as well as $r_0$ (see text).}
\label{fig_01}
\end{figure}

In the present study, we investigate the surface curvature effects on
collective excitations of electrons confined in one-dimensional (1D) dumbbell-shaped hollow nanotubes
(see Fig.~\ref{fig_01}).
This work is largely stimulated by the successful synthesis of 
peanut-shaped C$_{60}$ polymers \cite{onoe1,onoe2,onoe3,toda}.
It was discovered that under electron-beam radiation of a C$_{60}$ film,
C$_{60}$ molecules coalesce to form a peanut-shaped C$_{60}$ polymer being metallic \cite{beu}
and having a 1D hollow tubule structure
whose radius are periodically modulated along the tube axis.
Hence, the periodic variation in curvature
is expected to provide sizeable effects on quantum transport,
in which the system goes to Tomonaga-Luttinger liquid (TLL) states
due to the 1D nature \cite{voit,giamarchi}.
In fact, we shall see below that
periodic surface curvature in the dumbbell-like tube enhances the TLL exponent $\alpha$
describing the singularity of spectral functions,
which is qualitatively in agreement with the experimental results of 1D metallic
peanut-shaped C$_{60}$ polymers 
obtained using in situ photoelectron spectroscopy \cite{ito}.

\begin{figure}[ttt]
\begin{center}
\includegraphics[scale=0.45]{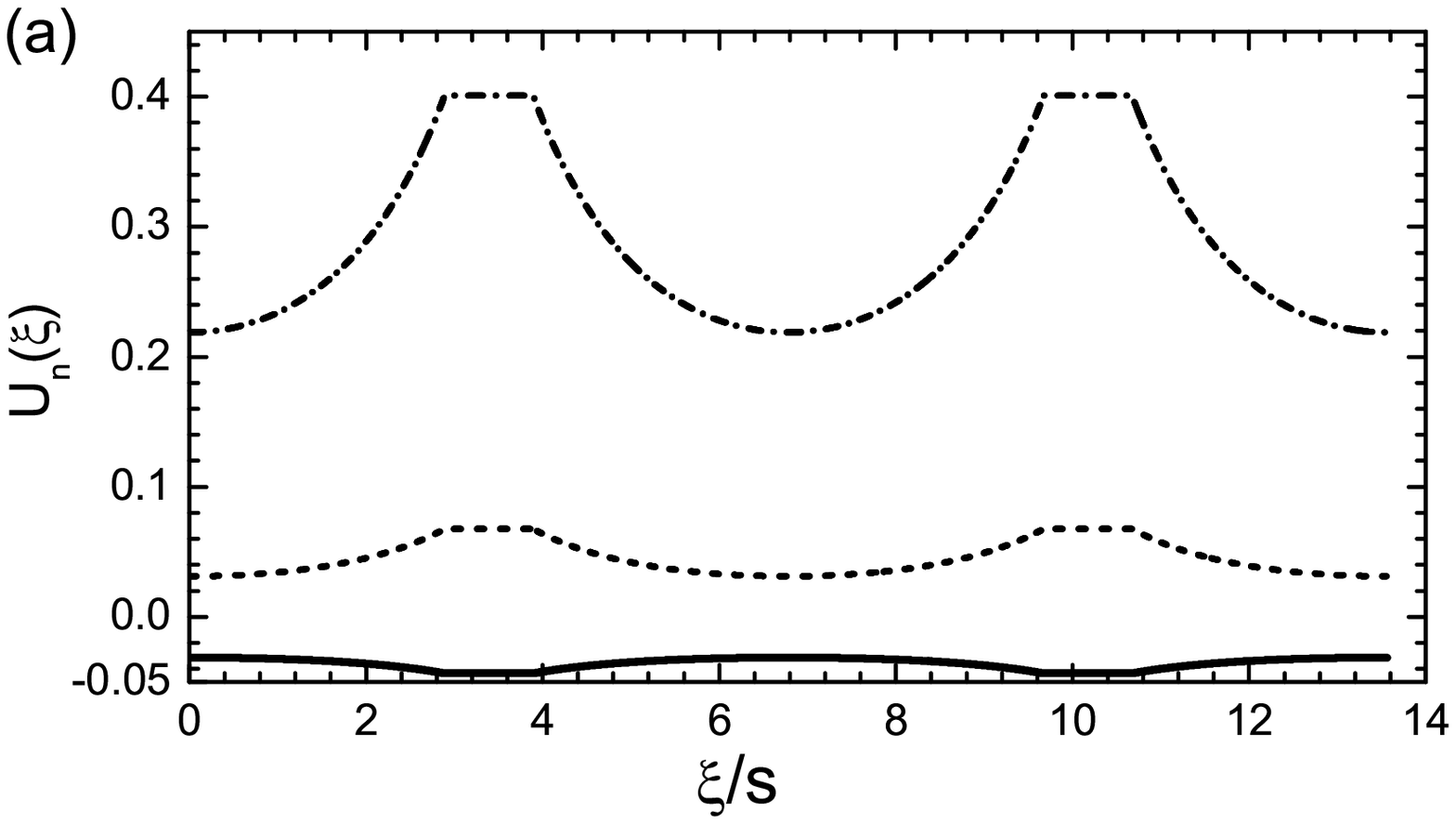}
\vspace*{-0.5cm}
\includegraphics[scale=0.45]{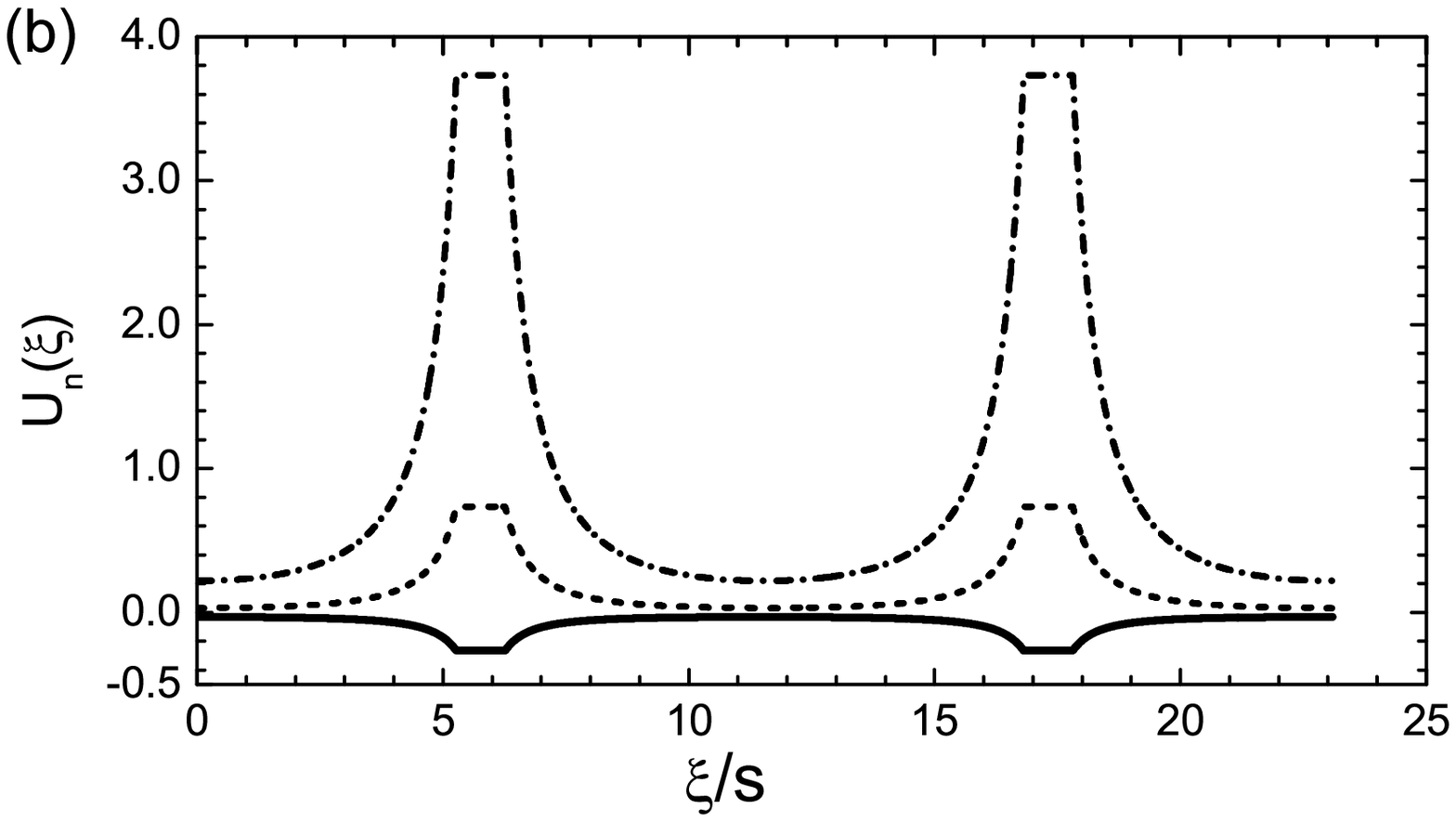}
\end{center}
\caption{Two-period profiles of the curvature-induced effective potential $U_n(\xi)$, 
where (a) $\delta r = 1.0$, $\Lambda = 6.78$ and
(b) $\delta r = 3.0$, $\Lambda = 11.5$ in units of $s$.
The abscissa $\xi$, defined by Eq.~(\ref{eq_03}), represents the line length
along the curve on the surface with a fixed $\theta$.
The integer $n$ characterizing the angular momentum of eigenstates
in the circumferential direction
ranges from $n=0$ (solid), $n=1$ (dashed) to $n=2$ (dashed-dotted).}
\label{fig_02}
\end{figure}

\section{Model and Method}

Figure \ref{fig_01} shows schematic illustration of a 1D dumbbell-shaped hollow nanotube.
It consists of an infinite chain of equi-separated spherical surfaces with radius $r_0$,
in which the chain is threaded by a thin hollow tube with radius $r_0 - \delta r$.
The neck length $s$ serves as the unit of length, and $r_0/s = 4$ is fixed throughout calculations
to mimic the actual geometry of peanut-shaped C$_{60}$ polymers.
Periodic modulation of the tube radius $r(z)$ along the $z$ direction 
is defined by 
\begin{equation}
r(z) = \left\{ 
\begin{array}{cl}
\sqrt{r_0^2 - z^2}; & |z|<\sigma \quad \mbox{(region I)} \\ [3pt]
r_0 - \delta r; & \sigma < z < \sigma + s \quad \mbox{(region II)}
\end{array}
\right.
\end{equation}
and $r(z) = r(z+\lambda)$, in which
\begin{equation}
\sigma = \sqrt{2r_0 \delta r - \delta r^2} \;\;\; \mbox{and} \;\;\; \lambda = s + 2\sigma.
\end{equation}

\begin{figure}[ttt]
\begin{center}
\includegraphics[scale=0.38]{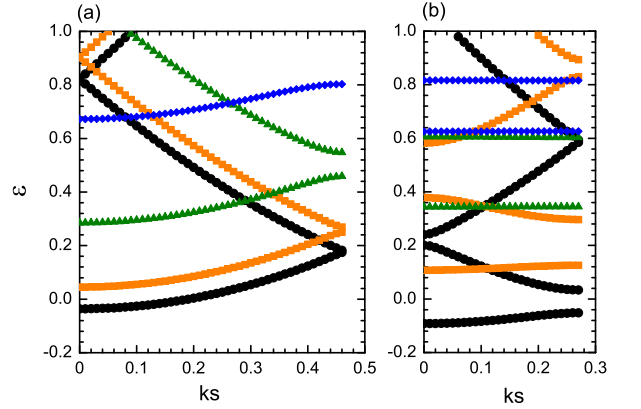}
\end{center}
\caption{(color online) Energy-band structures of dumbbell-shaped hollow cylinders
with (a) $\delta r = 1.0$ and (b) $\delta r = 3.0$ in units of $s$.
Branches belonging to the angular momentum index $n=0,1,2,3$ 
are represented by the symbols of circle, square, triangle, and diamond, respectively.}
\label{fig_03}
\end{figure}

We have established the Schr\"odinger equation $H\Psi = E \Psi$ for non-interacting spinless electrons
confined to the dumbbell surface
to deduce the TLL exponent $\alpha$ on the basis of the bozonization procedure \cite{voit,giamarchi}.
Because of the axial symmetry, single-particle eigenfunctions of the system
have the form of
$\Psi(z,\theta) = e^{in\theta} \psi_n(z)$.
We used the confining potential approach \cite{daCosta} incorporated with
the variable transformation from $z$ to $\xi$ defined by \cite{footnote}
\begin{equation}
\xi = \xi(z) = \int_0^z \sqrt{1 + \left(dr/dz \right)^2} dz',
\label{eq_03}
\end{equation}
or equivalently
\begin{equation}
\xi = \left\{
\begin{array}{ll}
r_0 \sin^{-1} (z/r_0) & \mbox{for region I}, \\ [2mm]
z + r_0 \sin^{-1} (\sigma/r_0) & \mbox{for region II}, 
\end{array}
\right.
\end{equation}
to obtain the differential equation for $\psi_n'(\xi) \equiv \sqrt{r(z)} \psi_n(z)$ 
such as \cite{fujita,shima}
\begin{equation}
\left[
- s^2 \frac{d^2}{d \xi^2} + U_n(\xi)
\right] \psi_n'(\xi) = \ep \psi_n'(\xi),
\;\;
\ep = \frac{2m^* s^2 E}{\hbar^2},
\label{eq_010}
\end{equation}
where
\begin{equation}
U_n (\xi) = \left( n^2 - \frac14 \right) \frac{s^2}{r(\xi)^2} - \frac{s^2}{4 r_0^2}.
\label{eq_06}
\end{equation}
Figure \ref{fig_02} shows the spatial profile of $U_n(\xi)$ for two periods, 
{\it i.e.,} for $0 < \xi < 2\Lambda$ with $\Lambda = \xi(\lambda)$.
At the neck region,
$U_n$ for $n=0$ takes minimum while those for $n\ge 1$ take maximum as understood from Eq.~(\ref{eq_06}).

Equation (\ref{eq_010}) is numerically solved by the Fourier expansion method;
see Ref.~\cite{shima} for details.
Figure \ref{fig_03} shows low-energy band structures
for $\delta r/s = 1.0$ in (a) and $\delta r/s = 3.0$ in (b), 
corresponding to those depicted in Fig.~\ref{fig_02}.
Energy gaps at the Brillouin zone boundary, $k = \pi/\Lambda$,
become wider for a larger $\delta r$,
as expected from the large amplitude of $|U_n(\xi)|$
with increasing $\delta r$.

\section{Results and Discussions}

We now consider the TLL states of dumbell-shaped tubes in which
the single-particle density of states $n(\omega)$
near the Fermi energy $E_F$ obeys the form \cite{voit,giamarchi}
\begin{equation}
n(\omega)\propto |\hbar \omega - E_F|^{\alpha}.
\label{eq_025}
\end{equation}
To make concise arguments, we assume $E_F$ to lie
in the lowest energy band.
We thus obtain\cite{voit,giamarchi}
\begin{equation}
\alpha = \frac{K+K^{-1}-2}{2},
\label{eq_030}
\end{equation}
and
\begin{equation}
K = \left[ \frac{2\pi \hbar v_F + V(2 k_F)}{2\pi \hbar v_F + 2 V(0) - V (2 k_F)} \right]^{1/2}.
\end{equation}
Here, $v_F = \hbar^{-1} dE/dk|_{k=k_F}$ is the Fermi velocity, and 
\begin{equation}
V(q) = - \frac{{\rm e}^2}{4\pi \varepsilon} \log \left[ ( q^2 + \kappa^2 ) r_0^2 \right]
\end{equation}
with the dielectric constant $\ep$ and the screening length $\kappa^{-1}$.
According to the bosonization procedure \cite{voit,giamarchi},
we set $\kappa s = 1.0\times 10^{-3}$ so as to be smaller than
all $k_F s$ values that we have chosen.
We also set the interaction-energy scale
${\rm e}^2/(4\pi \ep s) = 1.1 \times \hbar^2/(2 m^* s^2)$
by simulating that of C$_{60}$-related materials \cite{ep,mass}.

\begin{figure}[ttt]
\begin{center}
\includegraphics[scale=0.38]{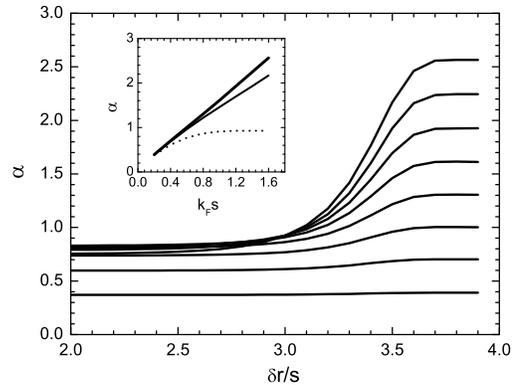}
\end{center}
\caption{$\delta r$-dependence of the TLL exponent $\alpha$. The Fermi wavenumber $k_F$
is increased from $k_F s = 0.2$ to $k_F s = 1.6$ from bottom.
Inset: $k_F$-dependence of $\alpha$ at $\delta r/s = 3.0$ (dashed), $3.5$ (thin), and $3.9$ (thick).}
\label{fig_04}
\end{figure}

Figure \ref{fig_04} shows the $\delta r$-dependence of $\alpha$
for different $k_F$ values.
Significant increases in $\alpha$ with $\delta r$ for $\delta r/a > 3.0$
are clearly observed, until reaching the plateau region at $\delta r > 3.7$.
These $\delta r$-driven shifts in $\alpha$ originate from the enhancement of 
the potential amplitude $|U_n(\xi)|$
that results in a monotonic decrease in $v_F$ with $\delta r$
at $\delta r/a > 3.0$.
The insets in Fig.~\ref{fig_04} shows the $k_F$-dependence of $\alpha$
at $\delta r/a=3.0, 3.5, 3.9$,
among which the last one presents an almost linear dependence on $k_F$.

The present results demonstrate that nonzero surface curvature 
yields diverse alterations in various kinds of power-law anomalies 
observed in TLL states,
such as the decay in Friedel oscillation \cite{TL3} 
and that of temperature (or voltage)-dependent conductance \cite{TL4},
where the power-law exponents depend on $\alpha$.
It is worthy to note that the similar increasing behavior of $\alpha$ was
observed in sinusoidal hollow tubes \cite{shima},
which implies that the presence of periodic curvature modulation rather than structural details
is important for the increase in $\alpha$ to occur.

\section{Conclusion}

In conclusion, we demonstrated the curvature-induced enhancement of the TLL exponent $\alpha$
in dumbbell-shaped hollow nanotubes.
The increase in $\alpha$ is attributed to 
the effective potential $U_n(\xi)$,
thus can be regarded as a geometric curvature effect on quantum transport
that is in the realm of existing materials of real peanut-shaped C$_{60}$ polymers.
We believe that experimental confirmation of the predictions 
will open a new field of science dealing with quantum electron systems on curved surfaces.

\section*{Acknowledgement}

We acknowledge K.~Yakubo, T.~Ito and Y.~Toda for stimulating discussions.
This study was supported by a Grant-in-Aid for Scientific Research 
on Innovative Areas and the one for Young Scientists (B) from the MEXT, Japan.
H.S is thankful for the financial supports from Kajima Foundation.
A part of the numerical simulations were carried out using
the facilities of the Supercomputer Center, ISSP, University of Tokyo.

%
%


\begin{thebibliography}{99}

\bibitem{daCosta} R.~C.~T.~da Costa, Phys.~Rev.~A {\bf 23} (1982) 1981.
\bibitem{DeWitt} B. De Witt, Phys.~Rev. {\bf 85} (1952) 635.
\bibitem{schuster} P.~C.~Schuster and R.~L.~Jaffe, Ann.~Phys. {\bf 307} (2003) 132.
\bibitem{prinz} V.~Ya Prinz, V.~A.~Seleznev, A.~K.~Gutakovsky, A.~V.~Chehovskiy, V.~V.~Preobrazhenskii, 
M.~A.~Putyato, and T.~A.~Gavrilova, Physica E (Amsterdam) {\bf 6} (2000) 828.
\bibitem{mci} D.~N.~McIlroy, A.~Alkhateeb, D.~Zhang, D.~E.~Aston, A.~C.~Marcy and M.~G.~Norton, 
J.~Phys.~Condens.~Matter {\bf 16} (2004) R415.
\bibitem{terrons} H.~Terrons and M.~Terrons, New J.~Phys. {\bf 5} (2003) 126.
\bibitem{twist1} I.~Arias and M.~Arroyo, Phys.~Rev.~Lett. 100, 085503 (2008).
\bibitem{ntn} H.~Shima and M.~Sato, Nanotechnology {\bf 19} (2008) 495705.
\bibitem{twist2} J.~Zou, X.~Huang, M.~Arroyo, and S.~L.~Zhang, J.~Appl.~Phys. {\bf 105} (2009) 033516.
\bibitem{pssa} H.~Shima and M.~Sato, phys.~stat.~sol.~(a) {\it in press}; DOI 10.1002/pssa.200881706.
\bibitem{exp1} G.~Cantele, D.~Ninno, and G.~Iadonisi, Phys.~Rev.~B {\bf 61} (2000) 13730.
\bibitem{exp2} H.~Aoki, M.~Koshino, D.~Takeda, H.~Morise, and K.~Kuroki, Phys.~Rev.~B {\bf 65} (2001) 035102.
\bibitem{exp3} D.~V.~Bulaev, V.~A.~Geyler, and V.~A.~Margulis, Phys.~Rev.~B {\bf 69} (2004) 195313.
\bibitem{exp4} A.~V.~Chaplik and R.~H.~Blick, New J.~Phys. {\bf 6} (2004) 33.
\bibitem{exp5} A.~Marchi, S.~Reggiani, M.~Rudan, and A.~Bertoni, Phys.~Rev.~B {\bf 72} (2005) 035403.
\bibitem{exp6} J.~Gravesen and M.~Willatzen, Phys.~Rev.~A {\bf 72} (2005) 032108.
\bibitem{exp7} H.~Taira and H.~Shima, Surf.~Sci. {\bf 601} (2007)	5270.
\bibitem{exp8} G.~Ferrari and G.~Cuoghi, Phys.~Rev.~Lett. {\bf 100} (2008) 230403.
\bibitem{exp9} V.~Atanasov, R.~Dandoloff, and A.~Saxena, Phys.~Rev.~B {\bf 79} (2009) 033404.
\bibitem{QHE} K.~J.~Friedland, A.~Siddiki, R.~Hey, H.~Kostial, A.~Riedel, and D.~K.~Maude,
Phys.~Rev.~B {\bf 79} (2009) 125320.
\bibitem{shota} S.~Ono and H.~Shima, Phys.~Rev.~B {\bf 79} (2009) 235407.
\bibitem{spin1} H.~Shima and Y.~Sakaniwa, J.~Phys.~A {\bf 39} (2006) 4921.
\bibitem{spin2} H.~Shima and Y.~Sakaniwa, J.~Stat.~Mech. (2006) P08017.
\bibitem{spin3} S.~K.~Baek, H.~Shima, and B.~J.~Kim, Phys.~Rev.~E {\bf 79} (2009) 060106.
\bibitem{spin4} S.~K.~Baek, P.~Minnhagen, H.~Shima, and B.~J.~Kim, Phys.~Rev.~E {\bf 80} (2009) 011133.
\bibitem{spin5} Y.~Sakaniwa and H.~Shima, Phys.~Rev.~E {\bf 80} (2009) 021103.
\bibitem{jensen} B.~Jensen, Phys.~Rev.~A {\bf 80} (2009) 022101.
\bibitem{taira} H.~Taira and H.~Shima, arXiv:0904.3149.
\bibitem{onoe1} J.~Onoe, T.~Nakayama, M.~Aono and T.~Hara, Appl.~Phys.~Lett. {\bf 82} (2003) 595.
\bibitem{onoe2} J.~Onoe, T.~Ito, S.~Kimura, K.~Ohno, Y.~Noguchi and S.~Ueda, Phys.~Rev.~B {\bf 75} (2007) 233410.
\bibitem{onoe3} J.~Onoe, T.~Ito, and S.~Kimura, J.~Appl.~Phys. {\bf 104} (2008) 103706.
\bibitem{toda} Y.~Toda, S.~Ryuzaki, and J.~Onoe, Appl.~Phys.~Lett. 92, 094102 (2008).
\bibitem{beu} T.~A.~Beu, J.~Onoe, and A.~Hida, Phys.~Rev.~B {\bf 72} (2005) 155416.
\bibitem{voit} J. Voit, Rep.~Prog.~Phys. {\bf 57} (1994) 977.
\bibitem{giamarchi} T.~Giamarchi, {\it Quantum Physics in One Dimension}, (Oxford University Press, 2004).
\bibitem{ito} T.~Ito, J.~Onoe, H.~Shima, H.~Yoshioka, and S.~I.~Kimura, {\it unpublished}.
\bibitem{footnote} The new variable $\xi$ corresponds to the line length
along the curve on the surface with a fixed $\theta$.
\bibitem{fujita} N.~Fujita, J.~Phys.~Soc.~Jpn, {\bf 73} (2004) 3115.
\bibitem{shima} H.~Shima, H.~Yoshioka and J.~Onoe, Phys.~Rev.~B {\bf 79} (2009) 201401(R).
\bibitem{ep} A.~F.~Hebard, R.~C.~Haddon, R.~M.~Fleming and A.~R.~Kortan, Appl.~Phys.~Lett. {\bf 59} (1991) 2109.
\bibitem{mass} A.~Oshiyama, S.~Saito, N.~Hamada and Y.~Miyamoto, J.~Phys.~Chem.~Solid., {\bf 53} (1992) 1457.
\bibitem{TL3} R.~Egger and H.~Grabert, Phys.~Rev.~Lett. {\bf 75} (1995) 3505.
\bibitem{TL4} C.~L.~Kane and M.~P.~A.~Fisher, Phys.~Rev.~B {\bf 46} (1992) 15233.
\end{thebibliography}
\end{document}